\documentclass[cits]{PoS}

\title{WMAP
 3-year polarization data: Implications for the reionization history}
\ShortTitle{WMAP 3-year polarization data:Implications for the reionization history}

\author{ {Lucia A. Popa}
\footnote{This work has been done in the framework of the
{\sc Planck} LFI activities.}\\

INAF/IASF Istituto Nazionale di Astrofisica,
Istituto di Astrofisica Spaziale e Fisica
Cosmica Bologna,
Via Gobetti 101, I-40129 Bologna, Italy\\
ISS Institute for Space Sciences, Bucharest-Magurele R-76900, Romania\\
E-mail: \email{popa@iasfbo.inaf.it lpopa@venus.nipne.ro} }

\abstract{
With the three-year data, the {\it Wilkinson Microwave Anisotropy Probe}
(WMAP3) produced a more
accurate determination of the electron
scattering optical depth,
downwarding its value from $\tau_{es}=0.17\pm 0.08$
obtained with the first-year data (WMAP1)
to $\tau_{es}=0.09\pm 0.03$.
As a consequence,
the structure formation in the $\Lambda$CDM best fit model
obtained WMAP3 is delayed relative to that of WMAP1. \\
We show that
the delay of structure formation
can not fully account for the reduction of $\tau_{es}$
from WMAP1 to WMAP3 when the radiative transfer
effects and feedback mechanisms
are took into account in computing the reionization history
of the Universe. \\
We also show that a PopIII stellar cluster with
a mass of $80 M_{\odot}$ and a heavy Larson initial mass function
has an ionizing efficiency high enough to account for
WMAP3 results, while in the case of WMAP1,
a higher stellar mass of $1000M_{\odot}$ was required.
}
\FullConference{}

\begin{document}
\section{Introduction}

With the three-year data on the anisotropy of the cosmic microwave background (CMB) and its polarization, the {\it Wilkinson Microwave Anisotropy Probe} (WMAP3) produced a more accurate determination of the electron
scattering optical depth, downwarding its value from $\tau_{es}=0.17\pm 0.08$ \cite{Spergel03}  obtained with the first-year data (WMAP1)
to $\tau_{es}=0.09\pm 0.03$ \cite{Page06, Spergel06},
consistent with an abrupt reionization at
redshift $z_{re} \simeq 11$, significantly later than $z_{re}\simeq 17$
as implied by WMPA1.

Common to most attemps to explain the high value of $\tau_{es}$ obtained by WMAP1
was the assumption that the globally-average ionization fraction  decreases over a
limited period of cosmic time (the so called ''double reionization``),
stimulating speculations regarding the efficiency of star formation and the escape fraction of the ionizing photons into the intergalactic medium (IGM).
Some studies \cite{Cen02, Wyithe04}
agree
on the necessity of the existence of a first generation of metal-free stars (PopIII stars) with heavy initial mass function (IMF).
Other works \cite{Ciardi03, Somerville00}
found that the metal enriched stars (PopII stars) are able to reionize the
Universe sufficiently early to produce the high values of $\tau_{es}$.
Other proposed physical mechanisms are radiative, the most important being  photoionization heating and  photodissociation cooling of the molecular hydrogen.
A recent work \cite{Furla04} critically examined
the plausibility of different  mechanisms
showing that the double reionization requires a rapid drop in
ionizing emissivity over a single recombination time that can be obtained
with unusual choices of the physical parameters.

Other most important changes of the cosmological parameters
from WMAP1 to WMAP3 are the reduction of the normalization of
the power spectrum on large scales ($\sigma_8=0.92\rightarrow0.76$)
and the reduction of the scalar spectral index of the primordial density
perturbations ($n_s=0.98\rightarrow0.74$).
As a consequence,
the structure formation in the $\Lambda$CDM model with
the primordial power spectrum of the density fluctuation obtained by WMAP3
is delayed relative to that of WMAP1.\\
Based on the simple assumption of constant ionizing efficiency,
a recent paper \cite{alvarez06} claims that the delay of structure formation controls the reionization in WMAP3 best fit model such that,
if ionizing efficiency is large enough to make reionization early and $\tau_{es}=0.17$ in WMAP1 case, the same efficiency is required
to have the reionization later and $\tau_{es}=0.09$
in WMAP3 case.

In this paper we show that
the delay of structure formation
can not fully account for the reduction of $\tau_{es}$ value
from WMAP1 to WMAP3 best fit models when the radiative transfer
effects and feedback mechanisms
are take into account in computing the reionization history
of the Universe. We also show that a PopIII stellar cluster with
a mass of $\sim 80 M_{\odot}$ and a heavy Larson IMF has
an ionizing efficiency high enough to account for the
value of $\tau_{es}$ obtained by WMAP3.\\
Throughout we assume a background cosmology consistent
with the most recent cosmological measurements \cite{Spergel06} with
energy density of $\Omega_m=0.24$ in matter, $\Omega_b=0.044$ in baryons,
$\Omega_{\Lambda}=0.76$ in cosmological constant, a Hubble constant of
$H_0$=72 km s$^{-1}$Mpc$^{-1}$ and adiabatic initial conditions
of the density fluctuations.

\section{WMAP 3-year data: implications for
the properties of ionizing sources}

\begin{table}[]
\begin{small}
\begin{center}
\begin{tabular}{cccccc}
\hline \hline
Model  & $n_s$&$\sigma_8$& $M/M_{\odot}$ & $\tau_{es}$   \\ \hline
WMAP1  & 0.99$\pm$0.04 & 0.92$\pm$0.1& $1000$ & 0.157$\pm$0.032 \\
WMAP3  & 0.961$\pm$0.017 &0.76$\pm$0.05&  $80$      & 0.093$\pm$0.012 \\
WMAP*& 0.99$\pm$0.04& 0.92$\pm$0.1&  $80$ & 0.130$\pm$0.032 \\ \hline
\end{tabular}
\end{center}
\end{small}
\caption[] { Model parameters}
\end{table}
\begin{figure}
\begin{center}
\includegraphics[width=12cm]{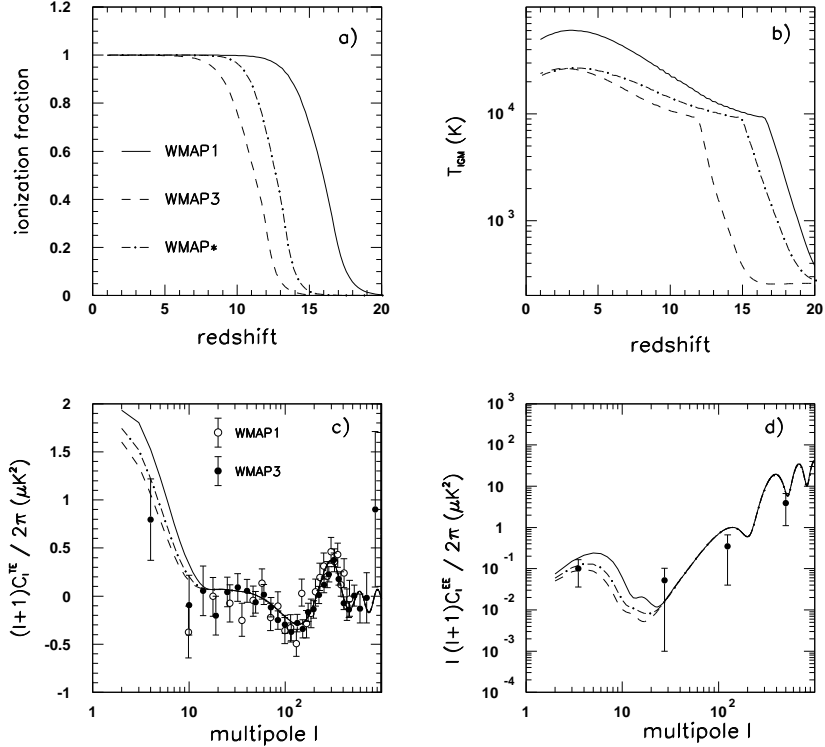}
\end{center}
\label{}
\vspace{-1cm}
\caption{Panel~a): Reionization histories obtained for the models
given in Table~1. Panel~b): Redshift evolution of the IGM
temperature corresponding to the reionization histories presented.
Panel~c): TE angular power spectra
for different reionization histories compared with WMAP1
and WMAP3 experimental measurements.
Panel~d): EE angular power spectra
for different reionization histories compared with WMAP3 experimental measurements.}
\end{figure}
The mean ultraviolet radiation background (UVB) intensity responsable for the cosmological reionization, $J(\nu_0,z_0)$, observed at the frequency
$\nu_0$ and the redshift $z_0$ produced by a population of sources characterized by the comoving emmisivity $\epsilon_{\nu}(z)$ can be written as:
\begin{eqnarray}
J(\nu_0,z_0)=\frac{c}{4\pi}
\int^{\infty}_{z_0} \epsilon_{\nu}(z) e^{-\tau_{eff}(\nu_0,z_0,z)}
\frac{dt}{dz}dz, \nonumber \\
\epsilon_{\nu}(z)={\bar{\it L}(z)} {\bar \tau}_{lf}f_{*}\frac{\omega_b}{\omega_m}
\frac{d}{dt} \int^{\infty}_{M_{min}(z)}\frac{dn}{dM_h}(M_h,z)M_hdM_h.
\end{eqnarray}
In the above equation $\nu=\nu_0(1+z)/(1+z_0)$, $(dt/dz)^{-1}=-H_0(1+z)\sqrt{\Omega_m(1+z)^3+\Omega_{\Lambda}}$
is the line element in our $\Lambda$CDM cosmological model,
$\tau_{eff}$ is the IGM effective otical depth,
$n(M_h,z)$ is the comoving number density of halos of mass $M_h$
at redshift $z$ given by Press-Shechter formalism, ${\bar{\it L}(z)}$
is the mean specific luminosity of the ionizing sources,
${\bar \tau}_{lf}$ is their mean lifetime and $f_{*}$ is the star formation efficiency.\\
We compute the mean specific luminosity ${\bar{\it L}(z)}$
for the emission of a PopIII stellar cluster according to \cite{Ferrara03}:
\begin{eqnarray}
{\bar{\it L}(z)}=\int_{M_{l}}^{M_{u}}F(\nu,M,z)\Phi(M)dM,
\hspace{0.4cm} F(\nu,M,z)=l^{st}_{\nu}(M)+l^{neb}_{\nu}(M)+l^{Ly\alpha}_{\nu}(M,z),
\end{eqnarray}
where: $\Phi(M)$ is the heavy Larson IMF,
$\Phi(M)\sim M^{-1}(M/M_{c})^{-1.35}$ with $M_c=15M_{\odot}$
normalized so that: $\int_{M_{l}}^{M_{u}}\Phi(M)dM=1$,
$l_{\nu}^{st}(M)$
is the spectrum of the star with the mass $M$, $l_{\nu}^{neb}(M)$
is the emission from its nebulae, $l_{\nu}^{Ly\alpha}(M,z)$
is the emission from $Ly\alpha$ photons and $M_{L}$ and $M_{u}$
are the lower and upper mass limits of the IMF.

We adopt the spectra of PopIII stars from \cite{Schaerer02}
with $M_l=80M_{\odot}$ and $M_u=1000M_{\odot}$ and
compute the redshift evolution of the reionization fraction for different
values of parameters $(n_s,\sigma_8,M)$ and the coresponding $\tau_{es}$
by using the equations (1.1) and (1.2).\\
Our computation includes all the radiative mechanisms
relevant for the primordial
gas dynamics: photo-ionization, photo-heating  and cooling of the
hydrogen and helium in the expanding Universe.
The mean UVB flux is obtained as  solution
to the radiative transfer equation
by assuming  a constant star formation efficiency $f_*=0.1$.\\
The model parameters and the corresponding values of $\tau_{es}$ are given in Table~1. The model WMAP* was constructed to have the same values for $n_s$ and $\sigma_8$ as WMAP1 and the same stellar mass as WMAP3.

We find that a stellar cluster with a mass of $M\simeq 80M_{\odot}$
has an ionizing efficiency high enough to account
for  WMAP3 value of $\tau_{es}$
while for the case of WMAP1 a higher stellar mass,
$M\simeq 1000M_{\odot}$, is needed.
For WMAP* model we obtain a value of the electron optical
depth of $\simeq0.13$ that can account for about $80\%$ from that obtained by WMAP1.\\
Panel a) from Figure~1 presents the redshift evolution
of the reionization fraction obtained for the models given in
Table~1. In Panel b) we show the redshift evolution of the IGM temperature
corresponding to the reionization histories presented and in
Panels c) and d) the corresponding  TE and EE polarization
angular power spectra compared with WMAP experimental measurements.\\

\section{Conclusion}

On the basis of these calculations,
we conclude that the delay of structure formation
in WMAP3 best fit model can not fully account for
the reduction of $\tau_{es}$ from WMAP1 to WMAP3 when
the radiative effects and feedback mechanisms  are take into account.\\
We find also that a PopIII stellar cluster with
a mass of $80 M_{\odot}$ and a heavy Larson IMF
can account for the WMAP3 results while, in the case of WMAP1,
a higher stellar mass of $1000M_{\odot}$ is required.

\end{document}